# Exploring inequality violations by classical hidden variables numerically

Sascha Vongehr

Institute of Materials Engineering, National Laboratory of Solid State Microstructures and College of Engineering and Applied Sciences, Nanjing University, Jiangsu, P. R. China

There are increasingly suggestions for computer simulations of quantum statistics which try to violate Bell type inequalities via classical, common cause correlations. The Clauser-Horne-Shimony-Holt (CHSH) inequality is very robust. However, we argue that with the Einstein-Podolsky-Rosen setup, the CHSH is inferior to the Bell inequality, although and because the latter must assume anti-correlation of entangled photon singlet states. We simulate how often quantum behavior violates both inequalities, depending on the number of photons. Violating Bell 99% of the time is argued to be an ideal benchmark. We present hidden variables that violate the Bell and CHSH inequalities with 50% probability, and ones which violate Bell 85% of the time when missing 13% anti-correlation. We discuss how to present the quantum correlations to a wide audience and conclude that, when defending against claims of hidden classicality, one should demand numerical simulations and insist on anti-correlation and the full amount of Bell violation. This preprint version adds a section on realisms and shows the actual programs with output.

## 1 Introduction

Quantum mechanics has real world applications, like quantum cryptography [1], which are based on quantum entanglement. Entanglement is proven to be non-classical by the experiments [2,3] and theory around the Einstein-Podolsky-Rosen (EPR) [4] paradox and John Bell's famous inequality [5,6]. While uncertainty and quantization *could* emerge from classical substrates, entanglement is incompatible with "local realism". Einstein-locality is here relativistic micro-causality (sometimes distinguished from Einstein-separability), and "realism" is in this context often ill-defined to mean 'counterfactual definiteness' (see added discussion of "realism" Section 6.1). Experimental violations of Bell inequalities disprove all non-contextual hidden variables. Variations of Bell's inequality, like the Clauser-Horne-Shimony-Holt (CHSH) inequality [7], have been strongly violated by experiment, most impressively with the closing of the 'communication loophole' [8]. Discussing exploitations of the still open 'detection loophole' is important for secure key distribution protocols [9,10], but nature cunningly deceiving us about it being classical would imply it wanting to do so rather than being a mere classical mechanism. Any fundamental detection loophole needs "further fact" uncertainties, like Bell's fifth position [11]. Nevertheless, computer simulations that seemingly violate Bell type inequalities via classical hidden variables are increasingly promoted and at times find their way into peer reviewed literature.

The derivation of the original Bell inequality (Section 2.3) assumes perfect anti-correlation of entangled photon measurements in the EPR setup. The CHSH inequality has the reputation of being superior for all purposes. It is harder to violate and its derivation does not need anti-correlation. Arguments that do not take advantage of the CHSH are frowned upon as being uninformed about the state of the art. This paper has developed out of exchanges with several researchers who found that numerical simulations of hidden variable models are not easily dismissed, and the models even gain further support from this circumstance. This is our main motivation; it is not interesting to simply relax an assumption in Bell's proof to see what happens. In fact, one of our conclusions is that we should *not* relax them. However, some of those who hold the CHSH to be far superior insist on that anti-correlation should be relaxed and that the CHSH must therefore be employed instead. This is a bad idea for various reasons, and all the different issues also in the discussion Section 4 were addressed in order to convince, successfully at last, a quantum information researcher that his elegant proof is inferior to EPR with Bell's inequality. Considering the growing need for dealing efficiently with proposed numerical simulations in this field, we establish a widely accessible understanding about what computer simulations can do and should be demanded to do. In the following, the simplest EPR experiment is introduced. A variation of the Bell proof is reviewed pedagogically, showing precisely how the inequality can be violated. A computer simulation of the true quantum statistics establishes the inequality violations' dependence on the number of simulated photons and benchmark parameters. We subsequently turn to simple programs that simulate classical hidden variables. We let them violate the Bell inequality and test whether the employed variables also violate the CHSH. A discussion of aspects that are almost all closely related to anti-correlation helps explaining why and how EPR, together with the Bell inequality, can be superior in the teaching and defense of quantum mechanics.

## 2  The EPR Setup and Inequalities

The EPR setup has a source of pairs of photons in its center. The photons are separated by sending them along the *x*-axis to Alice and Bob, who reside far away to the left and right, respectively; see [12] for a schematic illustration of the setup. Alice has a polarizing beam splitter (a calcite crystal) with two output channels. Alice's photon either exits channel "1", which leaves it horizontally polarized, or channel "0", which leads to vertical polarization (relative to the crystal's internal *z*-axis). The measurement is recorded as $A = 1$ or 0, respectively. Bob uses a similar beam splitter so that there are four possible measurement outcomes (*A*,*B*) for every photon pair: (0,0), (0,1), (1,0), or (1,1).

### 2.1  From Anti-Correlation to Sine-dependence

Every photon pair is prepared in 'singlet state entanglement', meaning that if the crystals are aligned in parallel, only the outcomes (0,1) and (1,0), for short U (for "Unequal"), will ever result. This is called anti-correlation. If the crystals are at an angle $\delta = (\beta - \alpha)$ relative to each other (rotated around the *x*-axis), outcomes depend on $\delta$. How do they depend on $\delta$? Anti-correlation implies that if $\delta = 0$ and Bob observes $B = 1$, Alice gets $A = 0$. Alice's photon *behaves as if* polarized orthogonally to Bob's measured one. The outcomes (0,0) and (1,1), for short E (for "Equal"), occur in the proportion $\sin^2(\delta)$. It is worthwhile to explain this dependence heuristically (see Section 4), because this very $\sin^2(\delta)$ is precisely what violates the Bell inequality.

### 2.2  The Inequality predicted by Quantum Mechanics

Every experiment starts with the preparation of a pair of photons. When they are about half-way on their paths to the crystals, Alice randomly rotates her crystal to let $\alpha = a\,\pi/8$ with *a* either 0 or 3. In other words, $a \in \{0, 3\}$ is selected at random by her, by throwing a coin or observing another quantum measurement. Bob adjusts his crystal similarly to $\beta = b\,\pi/8$ with $b \in \{0, 2\}$. No other angles need to be considered. The magnitudes of $\delta = (b - a)\,\pi/8$ are multiples of 22.5º, with $d = |b - a| \in \{0, \ldots, 3\}$. Hence, there are four equally likely cases: With $N_{\text{Total}} = 800$ photon pairs, the angles are expected to be about $N_d \approx 200$ times in each of the four configurations *d*. I avoid probabilities and consider only actual (never potential) counts $N$. The outcomes of all runs are counted by the $4*2 = 8$ counters $N_d(X)$, where $X \in \{E, U\}$. Anti-correlation leads to $N_0(E) = 0$ and $N_0(U) \approx 200$. Generally, it holds that

$$N_d(E) \approx N_d \sin^2(\delta), \qquad N_d(U) \approx N_d \cos^2(\delta). \qquad (1)$$

Apart from $N_0(E) = 0$, only three of these are important: $N_1(U) \approx 200 * \cos^2(-\pi/8) \approx 170$ alone is expected to be by 40 occurrences *larger* than the sum of $N_2(E) \approx 200 * \sin^2(\pi/4) \approx 100$ and the third number $N_3(U) \approx 200 * \cos^2(-3\pi/8) \approx 30$. We expect $N_1(U) > N_2(E) + N_3(U)$. A numerical simulation of 800 photon pairs (Section 3.1) leads on average only nine times out of 1000 runs to the coincidence of $N_1(U)$ being smaller than the right hand sum.

## 2.3 Hidden Variables and the Bell Inequality

Let us try to model the described experiment with help of hidden variables (HV). A pair of table tennis balls is prepared, say instructions are written on them, and then the pair is split, one ball going to the left to Alice, the other to Bob on the right. Before the balls arrive, Alice and Bob randomly select angles. Each ball results in a measurement 0 or 1 according to the angle it encounters and the HV (e.g. instructions) it carries. The instructions somehow prescribe "If $a = 0$, then $A = 0$", in short: "$A_0 = 0$". The ball at Bob's place cannot know how Alice has just adjusted her angle. She *might* have gotten $a = 0$. If so, Bob's measurement cannot be also 0 if he similarly has $b = 0$. Thus, the HV must prescribe the complementary information "$B_0 = 1$". If the HV prescribe $B_0$, prescribing $A_0$ explicitly is unnecessary, because anti-correlation equates $A_0$ and $(1 - B_0)$. Furthermore, only $A_3$ and $B_2$ must be somehow prescribed by the HV; otherwise, the occurrences $N_d(A,B)$ cannot reproduce the $\sin(\delta)$ dependence. In summary, the HV must contain the prescription of their degrees of freedom $(A_3, B_0, B_2)$. According to these three degrees of freedom, each pair of balls falls into only one of $2^3 = 8$ different classes, which we may index by $i = 4A_3 + 2B_0 + B_2$, so that $i$ is the result of taking $A_3B_0B_2$ as a binary number. For example, $N^2$ counts occurrences of $(0, 1, 0)$. The total number of pairs is $\sum_{i=0}^{7} N^i = 800$ again ($i$ is an index, not a power).

Let us briefly split a few hairs usually split by desperate believers in hidden classicality (one may skip this first): Alice's measurement of $A_0$ cannot change the value of $A_3$ in the past or that on Bob's ball. Her measurement of $A_0$ may change her ball's values; it does not matter, because if $A_0$ is measured, $A_3$ is not measured. $A_3$ is the value in case $a = 3$ *is* measured! We never assume any counterfactual definiteness that would not even be classically required: Hidden variables may randomly change. However, preparing HV, say (0, 0, 0), and then have them change with 30% probability to (0, 0, 1) on Bob's side in case $b = 2$, effectively means to prepare them three times out of ten as (0, 0, 1), not (0, 0, 0).

Let us proceed with the main derivation: Every pair encounters one of the four possible configurations of angles; hence, $N^i = N^i_0 + N^i_1 + N^i_2 + N^i_3$. All choices of angles occur about equally often and the HV cannot bias the choice (they have not arrived yet when the angles are chosen). Hence, all $N^i_d$ are expected to be roughly equal to $N^i/4$, which seems trivial but is *the* most important step, namely the very and only step where Einstein-locality comes in (closing the "communication loophole"; HV have not arrived when the angles are selected):

$$N^i_d \approx N^i/4 \qquad (2)$$

All the cases counted by $N^4_d$, $N^0_0$, $N^0_2$, $N^1_0$, $N^5_0$, $N^5_3$, and $N^6_1$ imply measurement outcome $(A,B) = (1,0)$. Equivalently, $N^0_1$, $N^0_3$, $N^1_3$, $N^2_1$, $N^2_2$, and $N^6_2$ correspond to (0,0), while $N^1_2$, $N^5_1$, $N^5_2$, $N^6_3$, $N^7_1$, and $N^7_3$ correspond to (1,1). Finally, $N^1_1$, $N^2_0$, $N^2_3$, $N^6_0$, $N^7_0$, $N^7_2$, and the four $N^3_d$ correspond to (0,1). This enumerates all the 32 possible $N^i_d$ exhaustively. Let us rearrange: All the cases $N^3_d$, $N^4_d$, $N^i_0$, $N^0_2$, $N^1_1$, $N^2_3$, $N^5_3$, $N^6_1$, and $N^7_2$ imply outcome (U), while $N^0_1$, $N^0_3$, $N^1_2$, $N^1_3$, $N^2_1$, $N^2_2$, $N^5_1$, $N^5_2$, $N^6_2$, $N^6_3$, $N^7_1$, and $N^7_3$ correspond to (E). In Section 2.2, the following three counters were important: $N_1(U) = N^1_1 + N^3_1 + N^4_1 + N^6_1 \approx (N^1+N^3+N^4+N^6)/4$, $N_2(E) \approx (N^1+N^2+N^5+N^6)/4$, and $N_3(U) \approx (N^2+N^3+N^4+N^5)/4$. Bell's inequality is here the mathematically trivial statement of that $N^1+N^3+N^4+N^6$ is by $2(N^2+N^5)$ *smaller* than $N^1+N^2+N^5+N^6$ and $N^2+N^3+N^4+N^5$ added together. In other words, it is expected that:

$$N_1(U) \leq N_2(E) + N_3(U) \qquad (3)$$

[ $N_3(E) > N_1(E) + N_2(U)$ ] is expected similarly but not necessary here.] Even if the HV are deliberately chosen in cunning ways, this inequality is expected because it derives from the randomness of the measurement angles leading to Eq.(2). Therefore, the described quantum experiment, where $N_1(U)$ alone is *larger* than the right hand sum by 40, cannot be modeled by any such HV.

## 2.4  Violating Inequalities

Our derivation of the original Bell proof allows us to see how to saturate the inequality: Not preparing any $i = 2$ or 5 pairs sets $N^2$ and $N^5$ equal to zero and thus ensures that the equal sign in Eq.(3) is expected. The random fluctuations around the equality then violate the Bell inequality Eq.(3) in half of all runs on average. Models that violate "often" and are presented as an advance toward a revolutionary discovery should be rejected by pointing out that HV which violate Bell 50% of the time while preserving anti-correlation are trivial.

### 2.4.1  *Allowing Missed Anti-Correlation*

The measurement procedures can try to "cheat" to get more than 50% violation. For example, if the hidden variables prescribe $i = 1$, then $B_2 = 1$ may increase $N^1_1$ or $N^1_2$. Alice can avoid the increase of $N^1_2$ by reporting $A_0 = 0$ (as if $i = 3$), but that increases $N_0(E)$ in case Bob reports $B_0 = 0$. Preserving anti-correlation requires Bob to collude with Alice: he must agree with her strategy in advance in order to report $B_0 = 1$ instead, but that would increase $N^3_3$ as often as $N^1_2$ is decreased, and so nothing is gained. Only by violating anti-correlation can they violate Bell more than half the time. This is similar for every other combination: At $i = 6$, Alice can decrease $N^6_2$ by misreporting $A_0 = 1$, but that increases $N_0(E)$ in case $b = 0$.

## 3  Computer Simulations

Computer modeling is attractive, because any 'locally realistic' model's behavior can in principle be realized by classical computers. That a computer can model a system is the essence of 'local realism': everything depends on locally present data; variables always have definite values even if they change randomly. Moreover, suggested classical models that are claimed to reproduce quantum behavior often carefully hide their wrong assumptions. Computer games where they must prove themselves practically have thus been discussed, see e.g. Vaidman's paper [13] and refs therein, for example in form of bets on the outcome of computer simulations [14]. The three-computer setup suggested by the "Quantum Randi Challenge" [15] itself constitutes a classical physical system rather than a simulation, so a Nobel Prize would be deserved for anybody who can provide Bell violating HV. Our computer simulations of the discussed EPR experiment as implemented in Mathematica™ consist each of less than ten lines of code. A slightly longer module (Fig. 2, Section 6.2) calculates the statistics, meaning to add up the various counters in the inequalities and tracking missed anti-correlation.

### 3.1  Calculating the Quantum Prediction Numerically

The program that calculates the true quantum behavior (Fig. 3, Section 6.2), simply enforces the quantum mechanical sine dependence discussed in Section 2.1 and does of course not use hidden variables. It prepares Alice's random angles $\alpha$ and random measurement outcomes $A$ with the following two lines of code:

```
Table[α[j] = If[Random[] < 0.5, 0, 3](π/8), {j, n}];
Table[A[j] = If[Random[] < 0.5, 0, 1], {j, n}];
```

Mathematica is an 'interpreted language' that stays very close to natural language; this allows to present the algorithm via explaining the code: "Random[]" produces a random number between zero and 1; thus, it is below 0.5 half of the time. If below 0.5, the entry after the first comma inside "If[…]" is returned, which is a zero in both lines. Otherwise, and thus also with probability of ½, the entry after the second comma inside "If[…]" is returned, which is a 3 in the first line, which multiplied with π/8 is the chosen Bell angle 3π/8 assigned to alpha. "Table[… , {j,n}]" repeats this until the photon pair counter $j$ has reached the desired total number $n$. The angles $\beta$ for Bob are prepared the same, but Bob's measurements $B$ are correlated with those of Alice in the way predicted by quantum mechanics: "B[j] = If[Random[] < (Sin[β[j] - α[j]])$^2$, A[j], 1 - A[j] ]". Alice's 0.5 (for her 50% probability) was replaced by the sin$^2$ dependence. $(1 - A)$ takes care of the anti-correlation, because $A$ and $B$ both only take the values 0 or 1.

This core algorithm has been run with total numbers of n = 40, 80, 160, 200, 400, and 800 simulated photon pairs. Quantum mechanics is expected to violate the Bell and CHSH inequalities, but since outcomes are random, sometimes the inequalities are obeyed. Running the program a thousand times for each total number of photon pairs resulted in the Bell inequality being obeyed 350, 253, 164, 123, 52, and 9 times respectively (Fig. 1). Considering always exactly 800 photon pairs is therefore sufficient didactically, because the program confirms the quantum mechanics prediction of that the Bell inequality is violated with a probability of around 99% ("almost always") already. Bigger numbers serve no purpose but to scare students and facilitate smokescreens around "cheats". The CHSH is more robust and was only obeyed 150, 103, 39, 21, 1, and 0 times, respectively. The CHSH is usually put into a form that results in a parameter $S$ which is classically expected to remain below 2, while quantum mechanics can lead to getting on average twice the square root of 2, i.e. about 2.828. Our angles have been chosen to maximize the Bell violation and result on average in an $S$ of about 2.4. Both inequalities' non-violation decreases approximately linearly with the logarithm of the number of photon pairs.

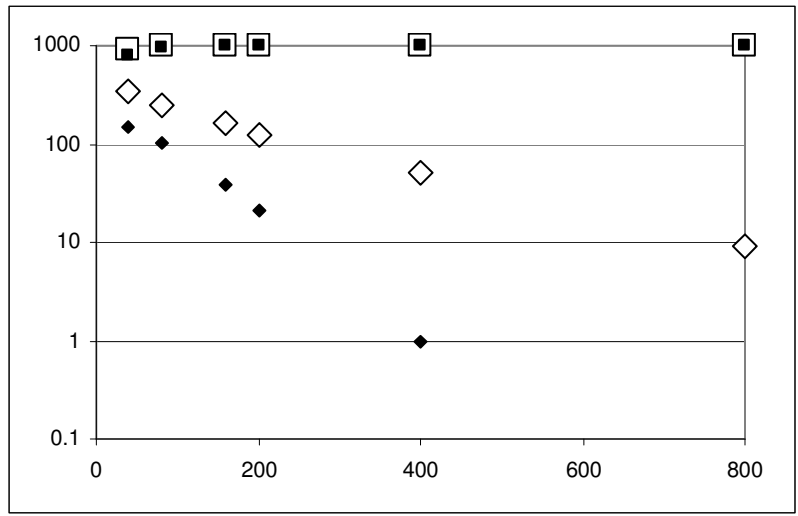

**Fig. 1:** Number of observed non-violations of Bell and CHSH inequalities (empty large and small filled symbols, respectively) out of 1000 trials when simulating the quantum behavior (diamond symbols) each time with a certain number of entangled photon pairs. Square shaped symbols show the behavior of the hidden variables considered by Bell.

## 3.2 Simulations with Hidden Variables

Having the number of photon pairs fixed to 800, we now look at the performance of HV models. Constructing Bell's random HV as discussed in Section 2.3 is accomplished by the line: "n = 800; Table[H[j, k] = If[Random[] < 0.5, 0, 1], {j, n}, {k, 3}]" (Fig. 4, Section 6.2). The counter $k$ goes to 3, so there are three random entries for every photon pair, meaning that the HV are conveniently the degrees of freedom ($A_3$, $B_0$, $B_2$) precisely as discussed. The measurement outcomes are now simply a read out of the HV picked by the angle setting, for example for Alice: "A[j] = If[α[j] == 0, 1 - H[j, 2], H[j, 1]]". For example, if the 736$^{th}$ α is zero, (1 – H[736, 2]) is assigned to $A$, where $k$ being equal to 2 returns the second entry in ($A_3$, $B_0$, $B_2$), i.e. the 736$^{th}$ $B_0$ value, and $A$ at zero α being equal to (1 – $B_0$) is indeed the correct and anti-correlated outcome. In case of 800 photon pairs, constructing HV, choosing measurement angles, and calculating the measurements as well as checking the correlation and adding counters for the evaluation of the Bell and CHSH statistics, are all accomplished in under one second on a ten year old PC. The unmodified random HV of the Bell proof did not once violate the inequalities during 1000 runs, which confirms that 800 pairs are sufficient in order to explore the topic. One has to go down to 160 photon pairs in order to start seeing inequality violation with about 1% probability. Running the program 1000 times, again for each total number of photon pairs as we did before with the simulation of the quantum statistics, resulted in the Bell inequality being obeyed 925, 983, 995, 1000, 1000, and 1000 times respectively (Fig. 1), and the CHSH being obeyed 794 and 969 times (note that this is less than Bell), and then 999, 1000, 1000, and 1000 times respectively.

Removing unwanted HV via the photon pair's index $i$ as discussed in Section 2.4 is done by first calculating that index ("i[j] = 4H[j, 1] + 2H[j, 2] + H[j, 3]"). Then the code "Table[H[j, k] = If[Or[i[j] == 2, i[j] == 5], 1 - H[j, 1], H[j, 1]], {j, n}, {k, 1}];" changes the HV to reflect a different $i$ whenever $i$ takes the undesired values 2 or 5. Now the program (Fig. 5, Section 6.2) indeed violates the Bell inequality 50% of the time, as expected from the proof (and not dependent on the number of photon pairs). What is not as easy to prove but seen immediately via the computer simulation is that the CHSH is also violated 50% of the time; therefore it provides no advantage here. Both inequalities are violated at random, meaning that many times, the CHSH is violated while the Bell inequality is obeyed.

A program (Fig. 6, Section 6.2) that modifies the HV according to Section 2.4.1, namely changes them at the moment of Alice's measurement as if Alice misreports $A_0 = 0$ in case $i = 1$ and $a = 0$, makes the model violate the Bell inequality about 85% of the time. Anti-correlation at equal angles decreased to 87% on average. The CHSH is still only violated 50% of the time. These HV are constructed to clearly show where the "cheating" is taking place. Some hidden variable theories have been shown to cleverly hide where they are cheating; they aim to maximize the magnitude of the violation, not clarity. The insistence on anti-correlation protects against such just as well as the insistence on the CHSH.

## 4 Discussion: Anti-Correlation versus CHSH

The CHSH works for quite mixed, not maximally entangled states that have no anti-correlation at any of the set angles. It is superior when analyzing the detection loophole in quantum cryptography. However, quantum research and its teaching should not revolve around whether loopholes may still not exclude spooky superdeterminism. Quantum mechanics has been confirmed via high energy particle physics, optics, and quantum chemistry. That theory predicts anti-correlation for singlet states. Most hidden variable models fail anti-correlation; another reason to insist on it. The theory does not predict that photons conspire to escape detection in just the right way to fool us. In this general view, the detection loophole is a technicality and will be narrowed by improving detectors, while only the communication loophole is a crucial issue, because Einstein-locality is crucial [16]. With the general attitude once more clarified, let us discuss neglected details.

### 4.1 Didactics of the Initial, Zero Angle Setup

Avoiding the $\delta = 0$ angle in order to maximize the CHSH violation makes the CHSH a bad choice for the teaching of quantum mechanics. The symmetric $\delta = 0$ scenario is an ideal starting point for several reasons. We can introduce almost the whole setup before other necessary angles (which are only two more in case of the Bell inequality) need to be mentioned. Its anti-correlation relates to well known optics: Although anti-correlation is difficult to explain with linear polarizations, the underlying reasons can be heuristically motivated with classical symmetries like angular momentum conservation. For example, if the photon-pair is prepared with zero overall rotation and Alice's photon is observed to be circularly polarized, say its electrical field vector rotates clockwise, Bob's photon must rotate the opposite way, because the total rotation is zero. Anti-correlation reflects the consistency of the behavior of photons with classical optics. This does not claim to derive quantum mechanics from classical physics. Photons are quantum. It must be stressed that we should never think of Alice's photon being actually flipped to a certain polarization direction, triggered by Bob's measurement! The light speed limit forbids any information from Bob's measurement to arrive at Alice's place in time for her measurement. Nevertheless, anti-correlation at $\delta = 0$ implies that Alice's photon will have behaved *as if* polarized orthogonal to Bob's measurement outcome. Only now must we mention non-zero angles, and we can do so via a powerful step in the argument, namely saying: However, $\delta$ may not even have been selected yet, say if Bob is further away from the photon source than Alice and if he delays his choice of angle sufficiently; therefore, anti-correlation at $\delta = 0$ implies that her photon will have behaved *as if* it was polarized orthogonally to Bob's photon at *all* angles $\delta$! This renders the photons' behavior equal to that known from polarizing sunglasses. Polarizing filters split the light's electrical field vector into orthogonal components. Simple geometry of projections (casting shadows) makes the components proportional to $\sin(\delta)$ and $\cos(\delta)$. Energy is proportional to the square of the field vectors, so energy is conserved: $\sin^2(\delta) + \cos^2(\delta) = 1$ (Pythagoras theorem). Students can even check this $\sin^2(\delta)$ dependence with a photo diode and a voltmeter. Energy is proportional to the number of photons. Hence, the light's photons must obey $\sin^2(\delta)$ and $\cos^2(\delta)$ as their probabilities for reaching the crystal's two output channels, or classical optics would not arise. It is worth to explain this heuristically, because the $\sin^2(\delta)$ is what violates the inequalities. Anti-correlation is merely $\sin(0) = 0$. The CHSH can hide all this, but we should not. The CHSH can be argued elegantly via logic of frustrated networks, but there are no comparatively 'physically intuitive' arguments.

### *4.1.1 Grasping Stricter than Complete Classical Correlation*

Extremely important for the understanding of the nature of quantum mechanics is Peres' statement that "*quantum phenomena are more disciplined*" [17] than even perfect classical common cause correlation can provide. The $\delta = 0$ situation's anti-correlation is preserved at all angles (recall "behaved *as if* it was polarized orthogonally to Bob's photon at *all* angles $\delta$" above). The classical, non-quantum correlation is fully preserved (not randomized), but obviously not the full issue. Defenders of hidden classicality are not alone in lacking this crucial insight into that the classical correlations they desire to re-introduce are still maximally present. Many-world/mind and modal interpretations can argue at this point that further correlations with "parallel branches" supply Peres' additionally necessary "discipline".

## 4.2  Amount of Violation, Convexity, Sampling, and arbitrary Thresholds

Some insist on the CHSH, because 95% reliable photon-pair emission, transmission, and coincidence-detection, which is sufficient for the CHSH to close the detection loophole, are conceivable, but perfect anti-correlation can neither be ensured by the photon-pair preparation nor by the angle settings. However, if we agree on that the Bell argument works in case of perfectly precise angles, it is not reasonable to argue that it suddenly fails if an angle misaligns by just 0.01 degrees, which destroys *perfect* anti-correlation. The *amount* of violation of the inequalities is too large for this to be an issue and it depends always on $\sin(\delta)$. We saw that even violating the CHSH 50% of the time is not sufficient to prove quantum behavior (Section 2.3); therefore, we need to consider the amount of violation anyway. Quantum mechanics violates the uncertainties 99% of the time (with 800 pairs) while keeping anti-correlation. Accepting quantum mechanics as far as confirmed demands the reproduction of its predictions for singlet states, which belong to the EPR setup as much as the specific angle settings. It does not matter if experiments will conceivably never prove more than 80 Bell violations out of 100 trials, perhaps because of technicalities or certain 'further fact' corrections (holographic black hole complementarity, Diosi-Penrose criterion, …), or the overall spin in the EPR setup's devices limiting angle resolution.

Bell's argument fails without the determinism ensured by anti-correlation. Relaxing anti-correlation must therefore discuss convexity and fair sampling. By convexity, every indeterministic hidden variable theory can be replaced by a deterministic one without affecting the observed statistics. This is implicitly argued in Bell's derivation, because the hidden variables can be much more complicated, while the proof and simulations need only look at the effective degrees of freedom of the hidden variables. Convexity is trivial: Anti-correlation at $\delta = 0$ is the full classical correlation, and any randomness via 'genuinely stochastic' hidden variables or imperfect anti-correlation can at most lead to less correlation, not "*more disciplined*" [17] correlation. The CHSH without anti-correlation does not entirely get around addressing convexity and fair sampling, but makes it difficult.

A related issue is that quantum mechanics *on principle* rejects statistical thresholds where a contender could claim to have disproved quantum mechanics empirically, because known quantum mechanics allows a challenger to win by sheer luck, no matter how small the probability, since the unitary (linear) state evolution does not allow a nonlinear cut-off on small probabilities. Winning bets in most futures is thus counterproductive. Instead, a small number of 800 photon pairs still obeys the Bell inequality relatively often. Such seemingly classical 'freak branches' belong to some of the most interesting questions at the foundations of quantum mechanics; none of this should be hidden in an honest defense.

## 5  Conclusion

The details of the original Bell proof provided a way to violate the Bell inequality via hidden variables. The numerical exploration violated also the CHSH 50% of the time, and worse by chance. Quantum mechanics gives no definite cutoff value that is safe to bet on. Programs simulating EPR are extremely simple. Proponents of hidden variable models should provide such. Both inequalities' *non*-violation by true quantum behavior decreases approximately linearly with the logarithm of the number of photons (Fig. 1). The quantum versus classical behavior is obvious with only 800 photon pairs (multiples of 80 are convenient in the Bell proof); large numbers only obscure. After discussing also the involved didactics and heuristics, it is recommended to insist on anti-correlation and to stick to the Bell inequality. We described how this presents the quantum correlations as additional constraints, over and above the present classical correlations, to a wide audience. Despite its elegance and robustness, arguments based around the CHSH are inferior in several aspects, especially in the EPR setup. The defense against claims of hidden classicality should demand the full extend of the quantum violation of the Bell inequality by simulations of the simple, three angles EPR setup. Betting that the CHSH will not be violated in any classical simulation is counterproductive.

# 6 Added Supplemental Material

## 6.1 Clarification of "Realism"

Relaxing "realism" triggers a wide spectrum of concerns from questioning personal identity (in how many worlds am "I") and responsible agency (due to modal totality being fixed by definition regardless my decisions) to fear of cultural relativism (does Everett relativity go too far?) and anti-science (description relativity being refused as similarly misunderstood 'postmodernism'). This has slowed the progress on quantum foundations and nourishes those who refuse it; hence, let us be careful. The ill-defining of "realism" in the physics literature, as well as the fact of "spooky" (A. Einstein) non-locality already contesting naïve realism [18,16], both suggest the term *direct* or *naïve* realism instead of the misleading "local realism". *Direct/naïve realism* [19] naively takes how things seem as if directly (without criticism) taken from the senses (there is also yet another use of "direct realism" in the over inflating philosophical literature). In physics, it supposes that objects with all their properties are a certain definite way '*really out there*', which *already includes* 'localism'. Some merely accept Everett relativity [20], some many-worlds [21,22] or minds [23,24] or different types of *modal* realisms (as for example popularized by Lewis [25]); others despise "parallel worlds", but all serious contenders know that *direct* realism is empirically excluded, period.

'*Empirical(ly based) modal realism*' may be a useful term to express that alternative situations ("parallel worlds") are labeled "real" because their correlations prove that they cannot be omitted from the physical description if it is to predict correctly. Since quantum mechanics allows empirical records that seem classical (freak branches), it empirically proves the limit of empirical verificationism. My '*tautological modal* *realism*' takes modal totality to be by definition the domain of theorizing of a theory of everything (ToE) possible, recognizing that the distinction between alternative worlds via "actualization" is fundamentally meaningless from the outset, even if actualization were not yet empirically proven to spread to multiple futures (e.g. by the inversion of Popper's proof [18,16]). This modal *realism* is thus '*anti-realism*', namely the refusal of verification transcendent distinctions (not to be confused with empirical verification).

### 6.2  Actual Programs and their Outputs (Screenshots)

**Figure 2**: The Statistical Analysis module

```
AC := {{N₀ = Σⱼ₌₁ⁿ If[α[j] == β[j], 1, 0], N_E0 = Σⱼ₌₁ⁿ If[And[α[j] == β[j], A[j] == B[j]], 1, 0]};
  If[N_E0 > 0, N[100 - 100 * N_E0 / N₀] "% Anti-correlation only. Model fails to describe the anti-correlation when Alice and Bob happen to measure with the same angle.",
   "Anti-correlation at equal angles OK."]};

BellT := {{N_U1 = Σⱼ₌₁ⁿ If[And[β[j] - α[j] == -π/8, A[j] ≠ B[j]], 1, 0], N_E2 = Σⱼ₌₁ⁿ If[And[β[j] - α[j] == π/4, A[j] == B[j]], 1, 0], N_U3 = Σⱼ₌₁ⁿ If[And[β[j] - α[j] == -3π/8, A[j] ≠ B[j]], 1, 0]},
  "The Bell inequality predicts that the first number is smaller than the sum of the second and third numbers.",
  If[N_U3 + N_E2 < N_U1, "Bell's inequality is violated!", "Bell inequality not violated."]};

CHSH := {{N₃ = Σⱼ₌₁ⁿ If[β[j] - α[j] == -3π/8, 1, 0], N_E3 = Σⱼ₌₁ⁿ If[And[β[j] - α[j] == -3π/8, A[j] == B[j]], 1, 0],
  N₁ = Σⱼ₌₁ⁿ If[β[j] - α[j] == -π/8, 1, 0], N_E1 = Σⱼ₌₁ⁿ If[And[β[j] - α[j] == -π/8, A[j] == B[j]], 1, 0], N₂ = Σⱼ₌₁ⁿ If[β[j] - α[j] == π/4, 1, 0],
  E₀ = 2 N_E0/N₀ - 1, E₁ = 2 N_E1/N₁ - 1, E₂ = 2 N_E2/N₂ - 1, E₃ = 2 N_E3/N₃ - 1}; CHV = N[Max[Abs[E₀ + E₁ + E₂ - E₃], Abs[E₀ + E₁ - E₂ + E₃], Abs[E₀ - E₁ + E₂ + E₃], Abs[E₁ + E₂ + E₃ - E₀]]],
  If[CHV > 2, "CHSH inequality is violated!", "CHSH inequality is not violated."]};
```

**Figure 3**: The simulation of quantum behavior needs knowledge of the relative angle. Below the green rectangle is the output of a typical run.

```
n = 800;
(*Random angles and random outcomes.*)
Table[α[j] = If[Random[] < 0.5, 0, 3] (π/8), {j, n}];
Table[A[j] = If[Random[] < 0.5, 0, 1], {j, n}];

(*Random angles also for Bob, but Bob's outcomes stay undetermined relative to Alice.*)
Table[β[j] = If[Random[] < 0.5, 0, 2] (π/8), {j, n}];

(*Bob's outcomes are correlated with Alice's angles.*)
Table[B[j] = If[Random[] < (Sin[β[j] - α[j]])^2, A[j], 1 - A[j]], {j, n}];

AC
MatrixForm[BellT]
CHSH
Clear[A, B, α, β]
```

{Anti-correlation at equal angles OK.}

$$\begin{pmatrix} \{145, 104, 21\} \\ \text{The Bell inequality predicts that the first number is smaller than the sum of the second and third numbers.} \\ \text{Bell's inequality is violated!} \end{pmatrix}$$

{2.42774, CHSH inequality is violated!}

**Figure 4**: The Bell hidden variables model. Below the green rectangle again the output of a typical run.

```
(*Hidden variables H*)
n = 800; Table[H[j, k] = If[Random[] < 0.5, 0, 1], {j, n}, {k, 3}];

(*Alice's angles α and measurements A*)
Table[α[j] = If[Random[] < 0.5, 0, 3] (π/8), {j, n}];
Table[A[j] = If[α[j] == 0, 1 - H[j, 2], H[j, 1]], {j, n}];

(*Bob's angles β and measurements B*)
Table[β[j] = If[Random[] < 0.5, 0, 2] (π/8), {j, n}];
Table[B[j] = H[j, If[β[j] == 0, 2, 3]], {j, n}];

(*Statistics*)
AC
MatrixForm[BellT]
CHSH
Clear[H, A, B, α, β]
```

{Anti-correlation at equal angles OK.}

$$\begin{pmatrix} \{115, 106, 96\} \\ \text{The Bell inequality predicts that the first number is smaller than the sum of the second and third numbers.} \\ \text{Bell inequality not violated.} \end{pmatrix}$$

{1.06351, CHSH inequality is not violated.}

**Figure 5**: These hidden variables violate the Bell and CHSH inequalities in 50% of all runs.

```
n = 800; Table[H[j, k] = If[Random[] < 0.5, 0, 1], {j, n}, {k, 3}];
(*The next formulas get rid of N^2 and N^5 and thus saturate the Bell inequality to make it
    an equality at high n. Bell is thus violated 50% of the time.*)
Table[i[j] = 4 H[j, 1] + 2 H[j, 2] + H[j, 3], {j, n}];
Table[H[j, k] = If[Or[i[j] == 2, i[j] == 5], 1 - H[j, 1], H[j, 1]], {j, n}, {k, 1}];

Table[α[j] = If[Random[] < 0.5, 0, 3] (π/8), {j, n}];
Table[A[j] = If[α[j] == 0, 1 - H[j, 2], H[j, 1]], {j, n}];

Table[β[j] = If[Random[] < 0.5, 0, 2] (π/8), {j, n}];
Table[B[j] = H[j, If[β[j] == 0, 2, 3]], {j, n}];

AC
MatrixForm[BellT]
CHSH
Clear[i, H, A, B, α, β]
```

{Anti-correlation at equal angles OK.}

$$\begin{pmatrix} \{142, 102, 40\} \\ \text{The Bell inequality predicts that the first number is smaller than the sum of the second and third numbers.} \\ \text{Bell inequality not violated.} \end{pmatrix}$$

{2.19492, CHSH inequality is violated!}

**Figure 6**: These hidden variables violate the Bell inequality 85% of the time (CHSH 50%) because they miss anti-correlation about 13% of the time.

```
n = 800; Table[H[j, k] = If[Random[] < 0.5, 0, 1], {j, n}, {k, 3}];
(*The next formulas get rid of N^2 and N^5 and thus saturate the Bell inequality to make it
    an equality at high n.*)
Table[i[j] = 4 H[j, 1] + 2 H[j, 2] + H[j, 3], {j, n}];
Table[H[j, k] = If[Or[i[j] == 2, i[j] == 5], 1 - H[j, 1], H[j, 1]], {j, n}, {k, 1}];

Table[α[j] = If[Random[] < 0.5, 0, 3] (π/8), {j, n}];
(*The next formula's red modification gets rid of N^1_2 and thus violates Anti-Correlation.*)
Table[A[j] = If[α[j] == 0, If[i[j] == 1, H[j, 2], 1 - H[j, 2]], H[j, 1]], {j, n}];

Table[β[j] = If[Random[] < 0.5, 0, 2] (π/8), {j, n}];
Table[B[j] = H[j, If[β[j] == 0, 2, 3]], {j, n}];

AC
MatrixForm[BellT]
CHSH
Clear[i, H, A, B, α, β]
```

{86.8293 % Anti-correlation only. Model fails to describe
    the anti-correlation when Alice and Bob happen to measure with the same angle.}

$$\begin{pmatrix} \{143, 75, 48\} \\ \text{The Bell inequality predicts that the first number is smaller than the sum of the second and third numbers.} \\ \text{Bell's inequality is violated!} \end{pmatrix}$$

{1.99165, CHSH inequality is not violated.}

[1] Ekert, A.K.: "Quantum cryptography based on Bell's theorem." Phys. Rev. **67**, 661 (1991)

[2] Aspect, A.; Grangier, P.; Roger, G.: "Experimental Tests of Realistic Local Theories via Bell's Theorem." Phys. Rev. Lett. **47**, 460-463 (1981)

[3] Aspect, A.; Grangier, P.; Roger, G.: "Experimental realization of Einstein-Podolsky-Rosen-Bohm Gedankenexperiment: a new violation of Bell's inequalities." Phys. Rev. Lett. **48**, 91-94 (1982)

[4] Einstein, A.; Podolsky, B.; Rosen, N.: "Can Quantum-Mechanical Description of Physical Reality be Considered Complete?" Phys. Rev. **47**(10), 777-780 (1935)

[5] Bell, J. S.: "On the Einstein Podolsky Rosen paradox." Physics **1(3)**, 195–200 (1964); reprinted in Bell, J.S. "Speakable and Unspeakable in Quantum Mechanics." 2nd ed., Cambridge: Cambridge University Press, 2004; S. M. Blinder: "Introduction to Quantum Mechanics." Amsterdam: Elsevier, 272-277 (2004)

[6] Bell, J.S.: "On the problem of hidden variables in quantum mechanics." Rev. Mod. Phys. **38**, 447–452 (1966)

[7] Clauser, J.F.; Horne, M.A.; Shimony, A.; Holt, R.A.: "Proposed Experiment to Test Local Hidden-Variables Theories." Phys. Rev. Lett. **23**, 880-884 (1969)

[8] Weihs, G.; Jennewein, T.; Simon, C.; Weinfurter, H.; Zeilinger, A.: "Violation of Bell's inequality under strict Einstein locality condition." Phys. Rev. Lett. **81**, 5039-5043 (1998)

[9] Barrett, J.; Hardy, L.; Kent, A.: "No Signaling and Quantum Key Distribution." Phys. Rev. Lett. **95**, 010503 (2005)

[10] Acin, A.; Gisin, N.; Masanes, L.: "From Bell's Theorem to Secure Quantum Key Distribution." Phys. Rev. Lett. **97**, 120405 (2006)

[11] Gill, R.D.: "Time, Finite Statistics, and Bell's Fifth Position." In *Proc of "Foundation of Probability and Physics – 2" Ser. Math. Modelling in Phys., Engin. And Cogn. Sci*, **5**, (2002) pp. 179-206. Vaxjo Univ. Press, (2003)

[12] Aspect, A.: "Bell's inequality test: more ideal than ever." Nature **394**,(18) 189-190 (1999)

[13] Vaidman, L.: "Test of Bell Inequalities." Phys. Lett. **A 286**, 241-244 (2001)

[14] Gill, R.D.; Larsson, J.A.: "Accardi contra Bell (cum mundi): The impossible coupling." In *Mathematical Statistics and Applications: Festschrift for Constance van Eeden*. Eds: M. Moore, S. Froda and C. L'eger, pp. 133-154. IMS Lecture Notes -- Monograph Series **42**, Institute of Mathematical Statistics, Beachwood, Ohio (2003)

[15] Vongehr, S.: Quantum Randi Challenge and Didactic Randi Challenges arxiv.org/abs/1207.5294v3 (2012)

[16] Vongehr, S.: "Historical Parallels between, and Modal Realism underlying Einstein and Everett Relativities." arxiv:1301.1972 [quant-ph] (2013)

[17] Peres, A.: "Unperformed Experiments have no Results" Am. J. Phys. **46**(7), 745-747 (1978); reprinted in: *Quantum Theory: Concepts and Methods*, Kluwer Academic Publishers, Dordrecht (1993), pp 161

[18] Vongehr, S.: "Realism escaping Wittgenstein's Silence: The Paradigm Shift that renders Quantum Mechanics Natural." 4th FQXi Essay Contest http://fqxi.org/community/forum/topic/1483, (2012)
18

[19] Stanford Encyclopedia of Philosophy: *Epistemological Problems of Perception*. http://plato.stanford.edu/entries/perception-episprob/#DirRea; Wikipedia: *Direct and Indirect Realism*. http://en.wikipedia.org/wiki/Direct_and_indirect_realism

[20] Everett, Hugh: "'Relative State' Formulation of Quantum Mechanics." Rev Mod Phys **29**, 454-462 (1957), reprinted in B. DeWitt and N. Graham (eds.), *The Many-Worlds Interpretation of Quantum Mechanics*, Princeton University Press (1973)

[21] B. DeWitt: "The Many-Universes Interpretation of Quantum Mechanics." in DeWitt, B. S., Graham, N. (eds.), *The Many-Worlds Interpretation of Quantum Mechanics*, Princeton University Press, Princeton NJ (1973)

[22] Deutsch, D.: "The Fabric of Reality." Allen Lane: New York (1997)

[23] D. Z. Albert, B. Loewer: "Interpreting the many-worlds interpretation." Synthese **77**, 195-213 (1988)

[24] M. Lockwood: ' "Many minds" interpretations of quantum mechanics.' Brit. J. Phil. Sci. **47**(2), 159-188 (1996)

[25] Lewis, D.K.: "On the Plurality of Worlds." Blackwell (1986)